\documentclass{article}
\usepackage{11lomcon}
\usepackage{txfonts}            
\usepackage{relsize}
\usepackage{graphicx}
\usepackage{epsfig}
\RequirePackage{xspace}
\def\Vub  {\ensuremath{|V_{ub}|}\xspace}
\def\Vcb  {\ensuremath{|V_{cb}|}\xspace}
\def\babar{\mbox{\slshape B\kern-0.1em{\smaller A}\kern-0.1em
    B\kern-0.1em{\smaller A\kern-0.2em R}}}
\def\invfb   {\ensuremath{\mbox{\,fb}^{-1}}\xspace}
\mathchardef\Upsilon="7107
\def\Y#1S{\ensuremath{\Upsilon{(#1S)}}\xspace}
\def\FourS   {\Y4S}
\def\nub     {\ensuremath{\overline{\nu}}\xspace}

\def\Bb      {\ensuremath{\Bbar}\xspace}
\def\Bu      {\ensuremath{B^+}\xspace}
\def\Bub     {\ensuremath{B^-}\xspace}
\def\piz     {\ensuremath{\pi^0}\xspace}
\def\KS      {\ensuremath{K^0_{\scriptscriptstyle S}}\xspace} 
 
\def\Dbar    {\kern 0.2em\overline{\kern -0.2em D}{}\xspace}

\def\mes     {\mbox{$m_{\rm ES}$}\xspace}

\def\BzBzb   {\ensuremath{\Bz {\kern -0.16em \Bzb}}\xspace}
\def\BpBm    {\ensuremath{\Bu {\kern -0.16em \Bub}}\xspace}
\def\Bz      {\ensuremath{B^0}\xspace}

\def\Bzb     {\ensuremath{\Bbar^0}\xspace}
\def\Dstarp  {\ensuremath{D^{*+}}\xspace}

\def\BR      {\ensuremath{\cal B}\xspace}
\def\Bbar    {\kern 0.18em\overline{\kern -0.18em B}{}\xspace}

\def\btoc    {\ensuremath{b \to c}}

\def\btou    {\ensuremath{b \to u}}
\def\btoulnu {\ensuremath{b \to u \ell \nu}}
\def\ra      {\ifmmode \rightarrow \else \ensuremath{\rightarrow}\xspace \fi}
\def\to      {\ifmmode \rightarrow \else \ensuremath{\rightarrow}\xspace \fi}
\def\Bxulnu  {\ifmmode \Bb \ra X_u \ell \bar{\nu} \else \ensuremath{\Bb \ra X_u \ell \bar{\nu}} \fi}
\def\Bxlnu   {\ifmmode \Bb \ra X \ell \bar{\nu} \else \ensuremath{\Bb \ra X \ell \bar{\nu}} \fi}

\newcommand {\Bxclnu}{\ensuremath{\Bb \ra X_c \ell \bar{\nu}}}

\newcommand {\mX}{\ensuremath{m_{X}}\hbox{ }}
\newcommand {\rusl}{\ensuremath{R_{u/sl}}}
\newcommand {\gevcc}{\ensuremath{{\mathrm{\,Ge\kern -0.1em V\!/}c^2}}\xspace}
\newcommand {\gevc}{\ensuremath{{\mathrm{\,Ge\kern -0.1em V\!/}c}}\xspace}
\newcommand {\mevcc}{\ensuremath{{\mathrm{\,Me\kern -0.1em V\!/}c^2}}\xspace}
\newcommand {\gev}{\ensuremath{\mathrm{\,Ge\kern -0.1em V}}\xspace}
\newcommand {\mev}{\ensuremath{\mathrm{\,Me\kern -0.1em V}}\xspace}
\newcommand {\mevc}{\ensuremath{{\mathrm{\,Me\kern -0.1em V\!/}c}}\xspace}

\newcommand {\bpiz} {\ensuremath{B^{\pm} \rightarrow \piz l \nu}\hbox{ }}
\newcommand {\bomega} {\ensuremath{B^{\pm} \rightarrow \omega l \nu}\hbox{ }}
\newcommand {\brhoz} {\ensuremath{B^{\pm} \rightarrow \rho^0 l \nu}\hbox{ }}

\newcommand {\gevtcf}{\ensuremath{{\mathrm{\,Ge\kern -0.1em V\! ^2 /}c^4}}\xspace}
\newcommand {\mmxn} {\ensuremath{\langle m_X^n \rangle}}

\newcommand{\gevn}{\ensuremath{{\mathrm{\,Ge\kern -0.1em V}^n}}\xspace}
\def\nubar    {\kern 0.18em\overline{\kern -0.18em \nu}{}\xspace}
\def\nulb     {\ensuremath{\nubar_\ell}\xspace}

\newcommand{\dsp}{\ensuremath{\Dstarp}\xspace}
\newcommand{\om}{\ifmmode{w} \else {$w$}\fi}
\newcommand{\omt} {\ifmmode {\tilde{w}} \else {$\tilde{w}$} \fi}
\newcommand{\psoft}{\ifmmode {{\pi_s}^+} \else {${\pi_s}^+$}\fi }

\newcommand{\ba}{\begin{eqnarray}}
\newcommand{\ea}{\end{eqnarray}}
\def\pip   {\ensuremath{\pi^+}\xspace}
\newcommand{\TBY}{\ifmmode{\theta_{\Bz, D^*\ell}} \else {$\theta_{\Bz, D^*\ell}$} \fi}
\newcommand{\BtoDss}{\mbox{$B\ra D^{*+} \pi \ell^- \nulb$}}
\newcommand{\dm}{\ifmmode {\Delta M} \else {$\Delta M$}\fi}

\newcommand{\hs}{\hspace}
\newcommand{\vs}{\vspace}
\newcommand{\Aone}{\ifmmode {{\cal A}_1} \else {${\cal A}_1$}\fi}
\newcommand{\rha}{\ifmmode{\mbox{\rho^2_{{\cal A}_1}}} \else {\mbox{$\rho^2_{{\cal A}_1}$}}\fi}

\begin{document}

\title{SEMILEPTONIC B DECAYS IN BABAR}

\author{A.Sarti \footnote{e-mail: asarti@slac.stanford.edu}}

\address{Dep. of Physics, The University of Ferrara and INFN, I-44100 Ferrara, Italy}

\maketitle\abstracts{~\babar~ measurements involving semileptonic decays 
of $B$ mesons are reviewed. Attention is focused on the extraction of 
\Vub~ and \Vcb~ elements of the Cabibbo-Kobayashi-Maskawa quark mixing 
matrix. Recent results of inclusive and exclusive approaches are presented.}

The study of semileptonic $B$ decays provides observables for
the extraction of \Vub~ and \Vcb~ elements of Cabibbo-Kobayashi-Maskawa 
matrix, the measurement of the $b$-quark mass and the extraction of 
non-perturbative QCD parameters ($\overline{\Lambda}$,$\lambda_{1}$).
The theoretical framework for such measurements consists
mainly of two different approaches: 
the Operator Product Expansion (OPE)[\cite{Chay:1990da}], 
used for the extraction of inclusive semileptonic rates, 
and the Heavy Quark Effective Theory, providing the tools for handling 
exclusive decays. Both approaches need corrections when used in 
real measurements: the analysis cuts, used to reject background events, 
are reducing the decay kinematic phase space and thus have to be
included in the OPE. The uncertainty on those corrections is currently 
giving the higher contribution to the systematic error.
Similarly, in \Vcb~ exclusive measurements, an 
extrapolation to a phase space boundary is needed when using HQET.\newline
From the experimental point of view the reconstruction of a large $B$
meson sample (BR(\btoulnu)$\sim 10^{-3}$) and an accurate reconstruction
of the $B$ decay chain (the current relative error on \Vcb~ is $\sim 2\%$)
are main issues. A dedicated $B$ meson reconstruction
technique has been set up in \babar~ to fulfill the following requirements.
A large sample of $B$ mesons can be collected by selecting hadronic 
decays $B_{reco} \ra  D^{(*)}X$, where $X$
represents a collection of hadrons composed of $\pi, K, \KS, \piz$ mesons.
The kinematic consistency of $B_{reco}$ candidates 
is checked with two variables, the beam energy-substituted mass 
$m_{ES} = \sqrt{s/4 - \vec{p}^{\,2}_B}$ and the energy difference 
$\Delta E = E_B - \sqrt{s}/2$. Here $\sqrt{s}$ is the total
energy in the $\Upsilon(4S)$ center of mass frame, and $\vec{p}_B$ and $E_B$
denote both the momentum and energy of the $B_{reco}$ candidate, in the same
frame.  The advantages of fully reconstructing one $B$ meson in the event are:
an easier subtraction of the continuum background, 
the knowledge of decay kinematics of the other $B$
(the only missing particle should be the neutrino)
and the possibility to request  flavor and charge 
correlations between the fully reconstructed $B$ meson and the one decaying 
semileptonically ($B_{recoil}$). On the recoil side, semileptonic events
can be selected using a cut on the lepton momentum 
($\sim 0.5\div1$\gevc) and on the neutrino four-momentum. 
The efficiency of this reconstruction technique is $\sim 0.1 \div 0.4\%$.

\section{Inclusive \Vub measurement}

\begin{figure}
\begin{centering}
\hskip -0.cm \epsfig{file=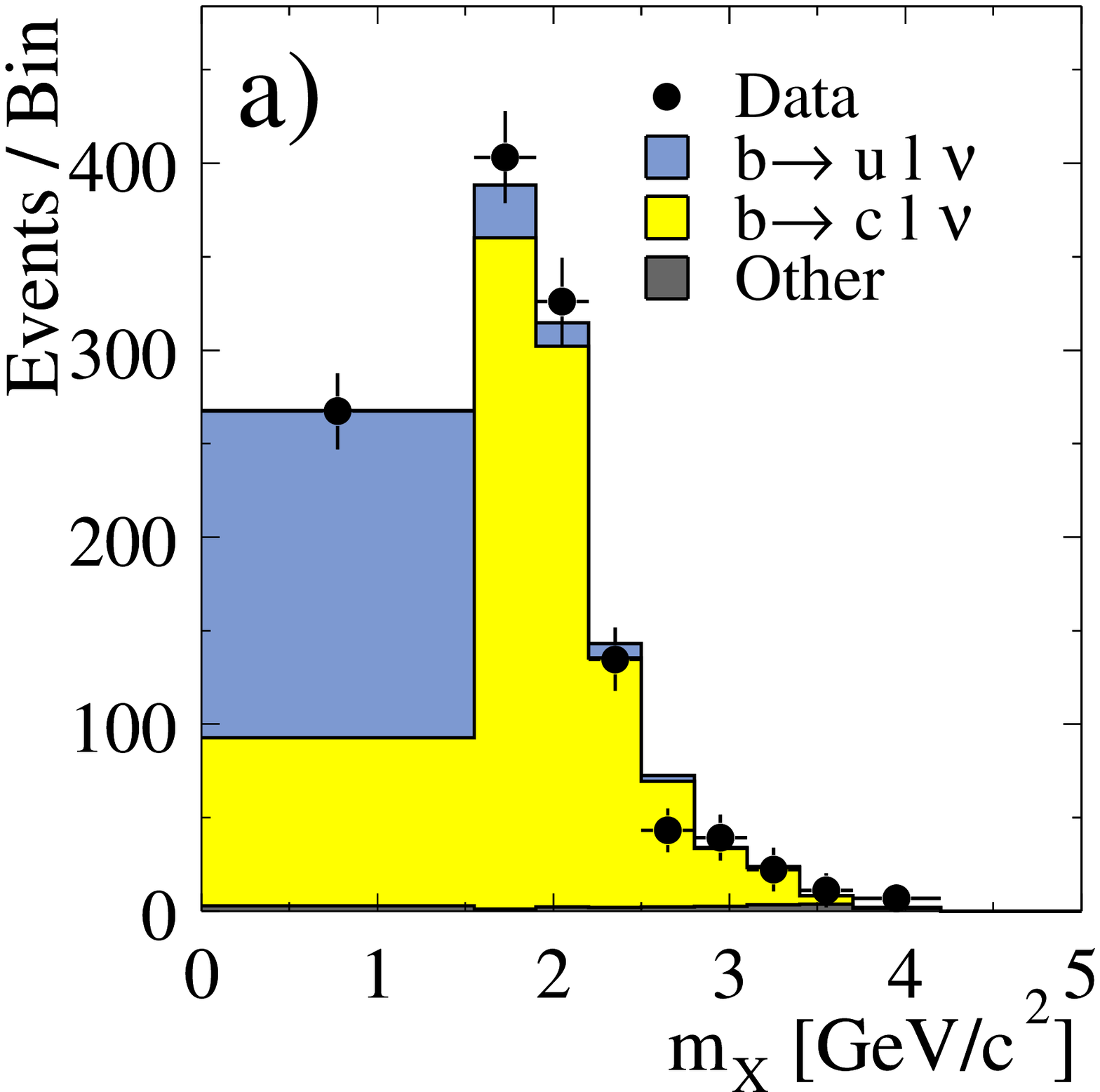,height=5.6cm} \hskip-0.3cm 
\epsfig{file=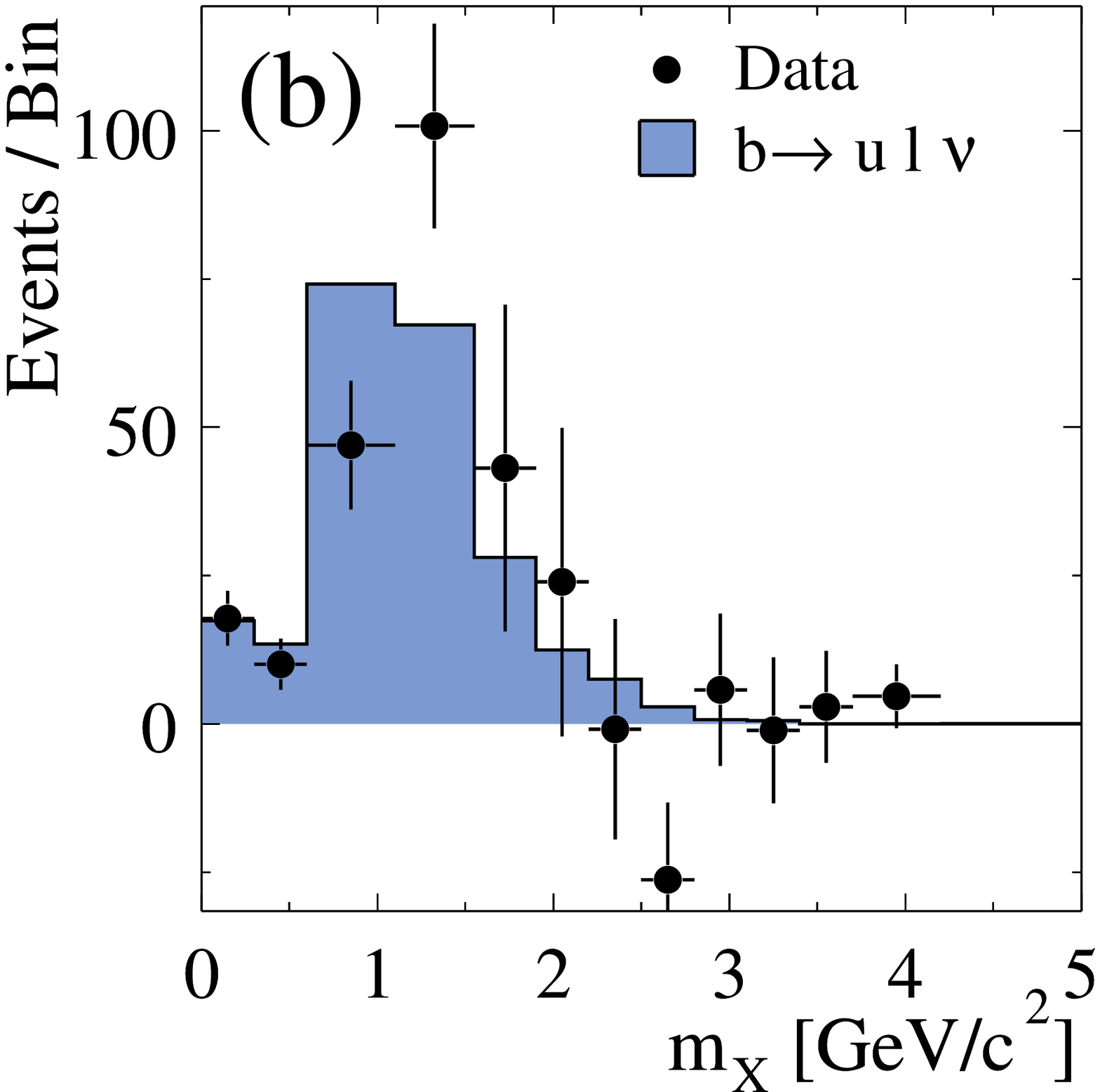,height=5.6cm} 
\caption{ The \mX distribution for \Bxlnu~ candidates for 
\babar~ inclusive analysis: a)
     data (points) and fit components, and b) 
     data and signal MC after subtraction of the $b\to c\ell\nu$ and the ``other'' backgrounds.
\label{fig:mxspectra}}
\end{centering}
\end{figure}

The main experimental issue of this analysis is the rejection of the huge
\Bxclnu~\footnote{Charge conjugation is implied.}
background (\BR(\Bxclnu)$\sim 60 \times $\BR(\Bxulnu)). 
The $B_{recoil}$ charmless semileptonic decays are selected with
a cut on the lepton momentum, a cut on the neutrino mass and a veto on
neutral and charged kaons. The residual background is determined 
from a fit of the hadronic invariant mass (\mX) distribution 
and \Vub~ is extracted from the measured semileptonic rate using the OPE
relation [\cite{pdg2002}]: $ \Vub =  0.00445 \sqrt{ (\frac{ B(\btoulnu) \cdot 1.55 ps } { 0.002 \cdot \tau_{B} } ) } \times (  1.0  \pm 0.020_{pert} \pm 0.052_{1/m_{b}^3} )$.
In order to reduce systematic uncertainties,
the ratio of branching ratios $\rusl$={\BR(\Bxulnu)/\BR(\Bxlnu)}
is determined from $N_u$, the observed number of \Bxulnu~ candidates 
with $\mX<1.55$\gevcc, and $N_{sl}$, the number of events with at 
least one charged lepton: 
\rusl$= \frac{N_u/(\varepsilon_{sel}^u \varepsilon_{\mX}^u)}{N_{sl}}  \times \frac{\varepsilon_l^{sl} \varepsilon_{reco}^{sl} } {\varepsilon_l^u \varepsilon_{reco}^u }$.
Here  $\varepsilon^u_{sel}$ is the efficiency for selecting $\Bxulnu$ decays
once a $\Bxlnu$ candidate has been identified, $\varepsilon^u_{\mX}$ is the
fraction of signal events with $\mX < 1.55$\gevcc, 
$\varepsilon_l^{sl}/\varepsilon_l^u$ corrects for the
difference in the efficiency of the lepton momentum cut for $\Bxlnu$ and 
$\Bxulnu$ decays, and $\varepsilon_{reco}^{sl}/\varepsilon_{reco}^u$
accounts for a possible efficiency difference in the $B_{reco}$
reconstruction in events with $\Bxlnu$ and $\Bxulnu$ decays.
$N_{sl}$ is derived from a fit to the \mes distribution.
$N_u$ is extracted from the $\mX$ distribution by a minimum $\chi^2$
fit to the sum of three contributions: the signal, the background
$N_{c}$ from $\Bxclnu$, and a background of $<1\%$ from other sources 
(misidentified leptons, secondary $\tau$ and charm decays). 
Fig.~\ref{fig:mxspectra}a shows the fitted $\mX$ distribution.
Fig.~\ref{fig:mxspectra}b shows the $\mX$ distribution
after background subtraction.
By using 82\invfb integrated luminosity on the $\Upsilon(4S)$ peak,
\babar~ obtains $\Vub~ = (4.62 \pm 0.28 \pm 0.27 \pm 0.40 \pm 0.26)
\times 10^{-3}$, where errors are respectively 
statistical, detector systematic,  theoretical model and 
propagation of error from the OPE relation [\cite{pdg2002}]. 
The signal over background ratio (S/B) is $~$1.7 
(higher than any previous inclusive analysis) 
and the main error comes from the parametrization of the 
$b$-quark Fermi motion inside the $B$ meson parametrization. \newline
Reduction of theoretical systematic error, related to the shape function
parameterization of the Fermi motion, can be achieved by adding a cut 
on the di-lepton pair invariant mass ($q^2$), to the \mX~ one, which  
allows for a reduction of systematic error of $\sim35\%$ on \Vub .

\section{Exclusive \Vub~ measurement}

\begin{figure}
 \begin{centering}
 \epsfig{file=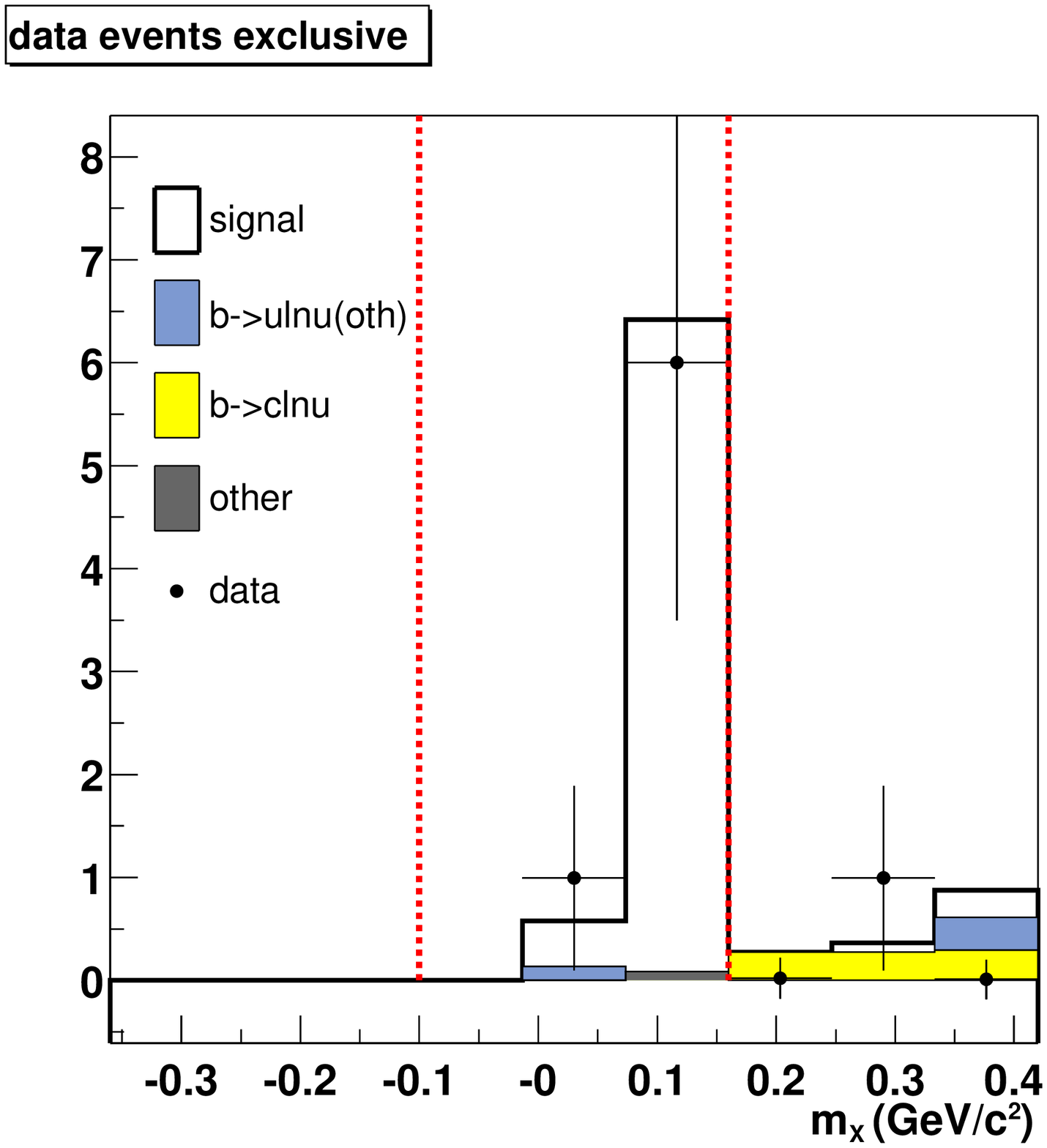,width=3.5cm}
 \epsfig{file=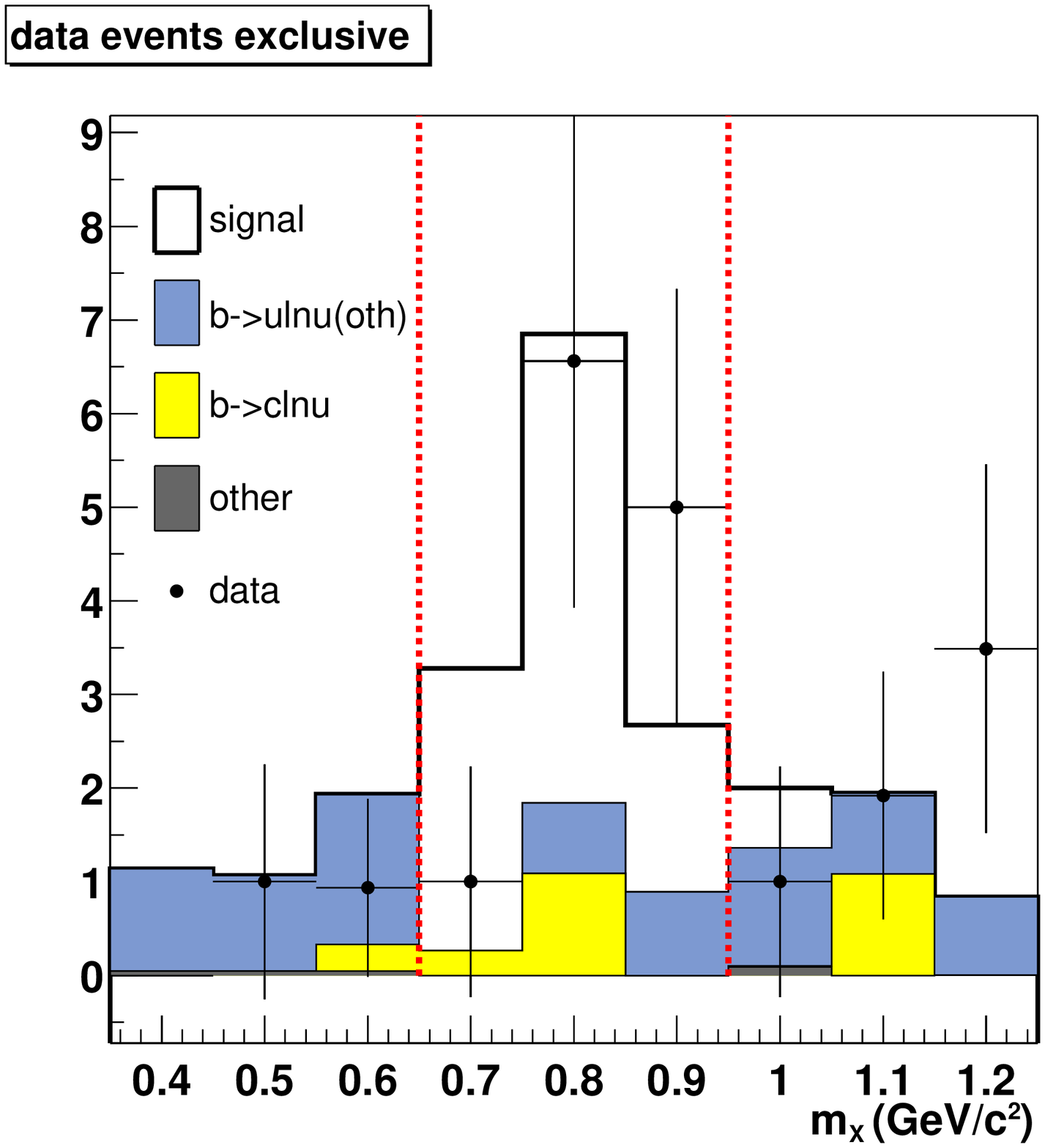,width=3.5cm}
 \epsfig{file=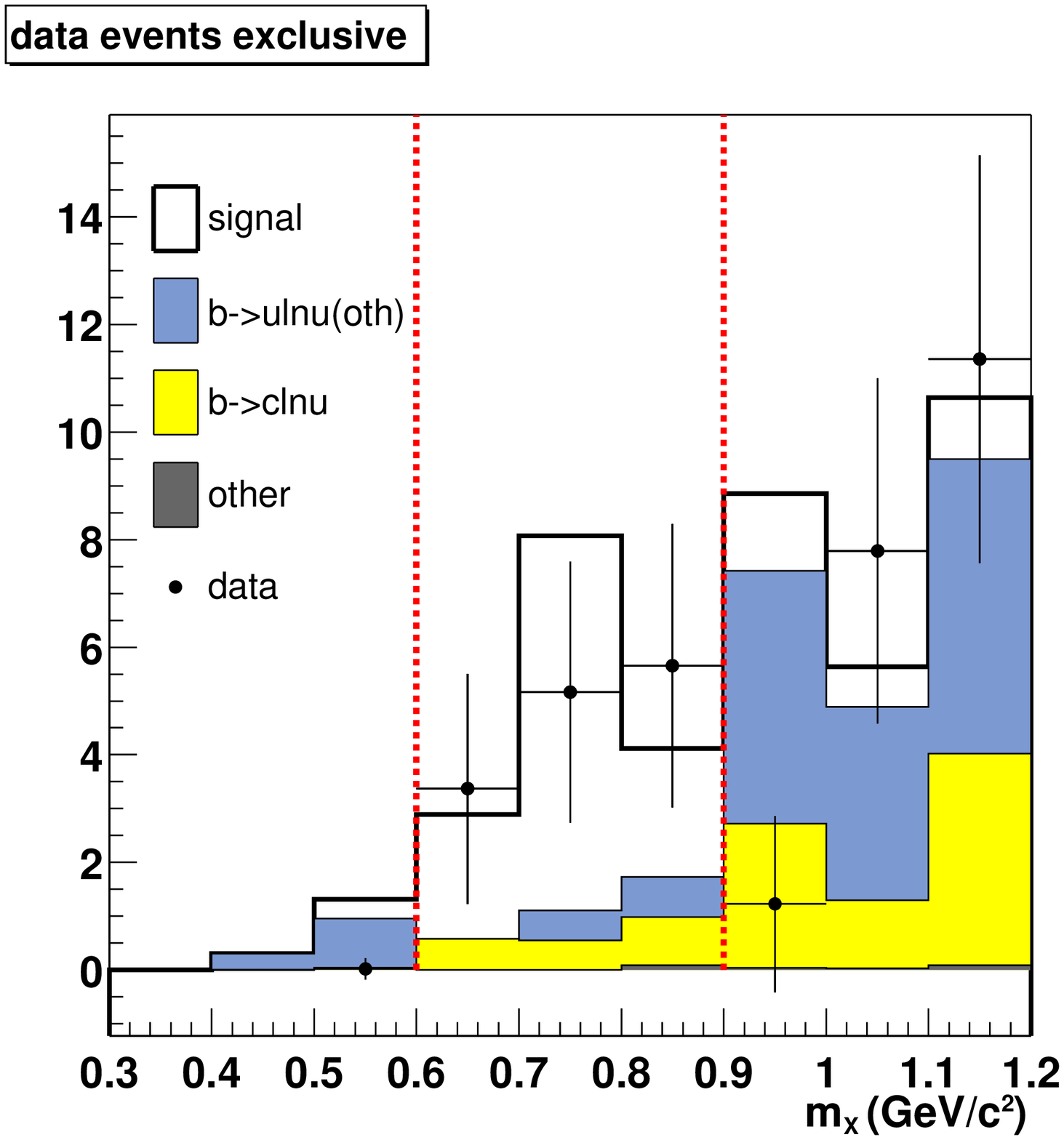,width=3.5cm}
 \caption{Projections in the \mX variable for the \babar~ \Vub~ exclusive analysis.
Vertical dotted lines represent the signal region. Left: \bpiz; middle: \brhoz; right: \bomega.
 \label{fig:fitall}}
 \end{centering}
\end{figure}

Using the same $B$ meson reconstruction technique as in the inclusive 
analysis, and with a similar analysis strategy and signal events 
selection, it is possible to study  the exclusive \Bxulnu~ decays.
The high purity of the sample of reconstructed $B$ mesons 
makes possible to use the mass of the hadronic system (\mX) to separate 
the resonances. \bpiz, \brhoz~ and \bomega ~ decays are selected applying 
constraints on the missing mass and \mX .  The measured exclusive branching ratios are:
\BR(\bpiz)$ = (0.78 \pm  0.32_{stat} \pm 0.13_{syst}) 10^{-4}$, 
\BR(\brhoz)$ = (0.99 \pm  0.37_{stat} \pm 0.19_{syst}) 10^{-4}$ and
\BR(\bomega)$ = (2.20 \pm  0.92_{stat} \pm 0.57_{syst}) 10^{-4}$,
where the main contribution to the systematic error comes from the 
uncertainty on signal MC modeling and the fit to the \mes distributions 
used to extract the number of signal events. The \BR(\brhoz) result
is obtained applying a cut on the two pions invariant mass
(0.65\gevcc $\le m_{\pip\pi^{-}} \le$0.95\gevcc).
In Figure ~\ref{fig:fitall} the projection of the results on the \mX variable
is shown.

\section{Inclusive \Vcb~ measurement}

\begin{figure}
  \begin{centering}
    \epsfig{file=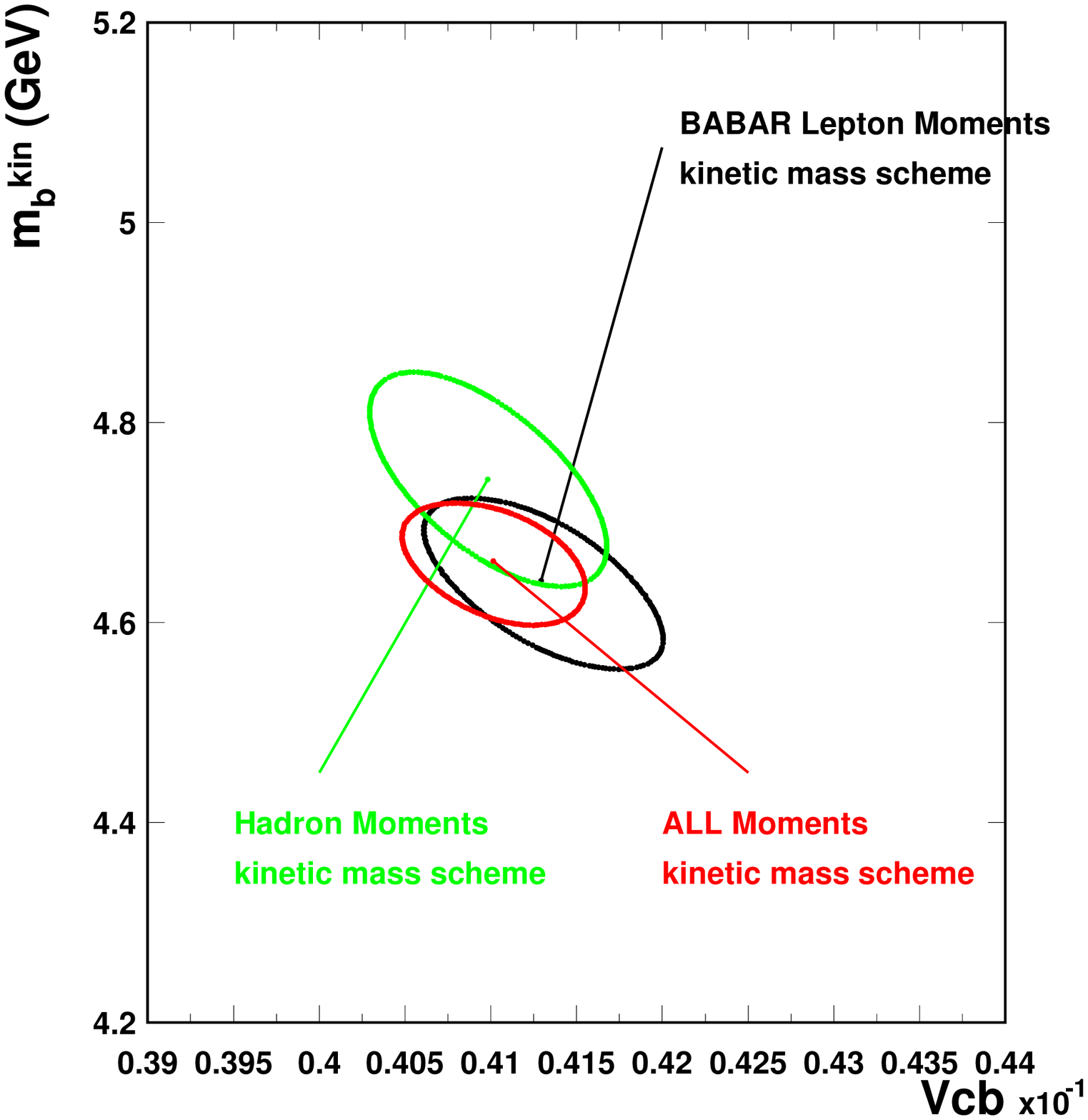,height=6.0cm}
    \epsfig{file=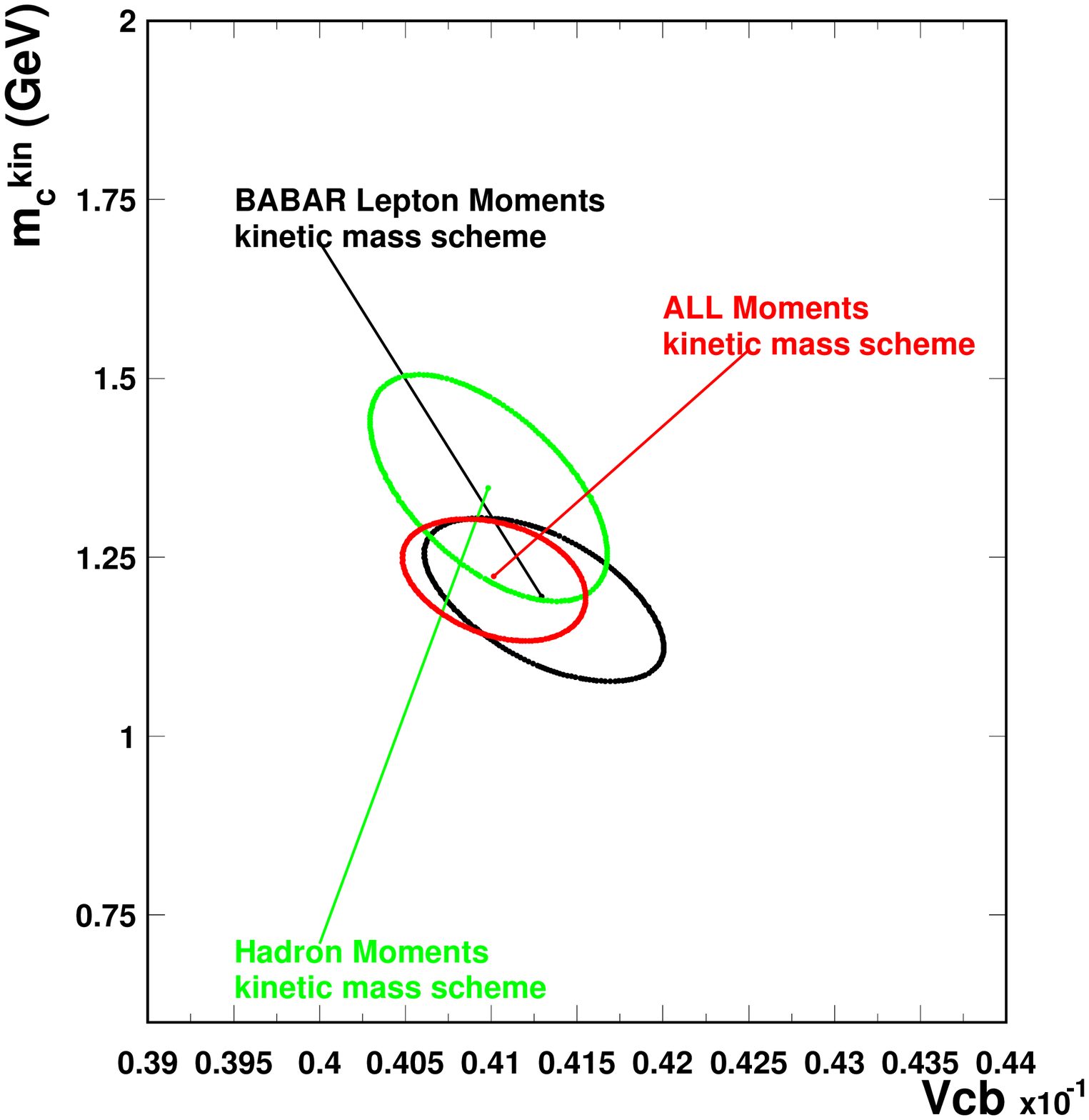,height=6.0cm}
    \caption{Fit results with contours corresponding to 
$\Delta \chi^2=1$ for (left) \Vcb~ versus $m_b$, and (right)
\Vcb~ versus $m_c$, separately for 
fits using the hadron mass, the lepton energy, and all moments.
    \label{fig:ellipses}}
   \end{centering}
  \vspace{-0.3cm}
\end{figure}

By measuring the first few moments \mmxn~ of the hadronic mass distributions 
in $\Bb\to X_c \ell^-\nub$ decays, 
it is possible to constrain the OPE parameters and extract \Vcb~ and the 
heavy quark masses $m_b$ and $m_c$. The measurement of the 
lepton energy moments can also improve the determination 
of \Vcb ~[\cite{interpret}]. $B$ mesons have been fully reconstructed with 
the technique discussed above. The semileptonic events selection proceeds
using a cut on the lepton momentum and on the missing mass. Charge and
flavor correlations are requested.  
The main sources of systematic errors are the precision in the 
modeling of the detector efficiency and particle reconstruction, the 
subtraction of the combinatorial background  of the $B_{\rm reco}$ sample, 
the residual background estimate and the uncertainties  in the modeling of 
the hadronic states.  The results do not depend on assumptions for branching 
fractions and mass distributions for higher mass hadronic states. 
Using a Heavy Quark Expansion in the kinetic mass scheme to order $1/m_b^3$, 
we extract the branching fraction, ${\cal B}_{c\ell\nu}=(10.62 \pm 0.16_{exp} 
\pm 0.06_{HQE}) \% $, and  the CKM matrix element, $|V_{cb}|= (41.25 \pm 
0.45_{exp} \pm  0.41_{HQE} \pm 0.62_{theory})\times 10^{-3}$, with 
significantly reduced uncertainties.  
Figures~\ref{fig:ellipses}(left) and ~\ref{fig:ellipses}(right) are showing
the $\Delta \chi^2=1$ ellipses for $|V_{cb}|$ versus $m_b$ and $|V_{cb}|$ 
versus $m_c$ for the standard fit to all data and separate 
fits of the hadron and lepton moments, but including the truncated 
branching fractions in both. 

\section{Exclusive \Vcb~ measurement}

\begin{figure}
\begin{center}
\begin{tabular}{ll}
\hs{-0.7cm}\includegraphics[height=5.5cm,width=5.5cm]{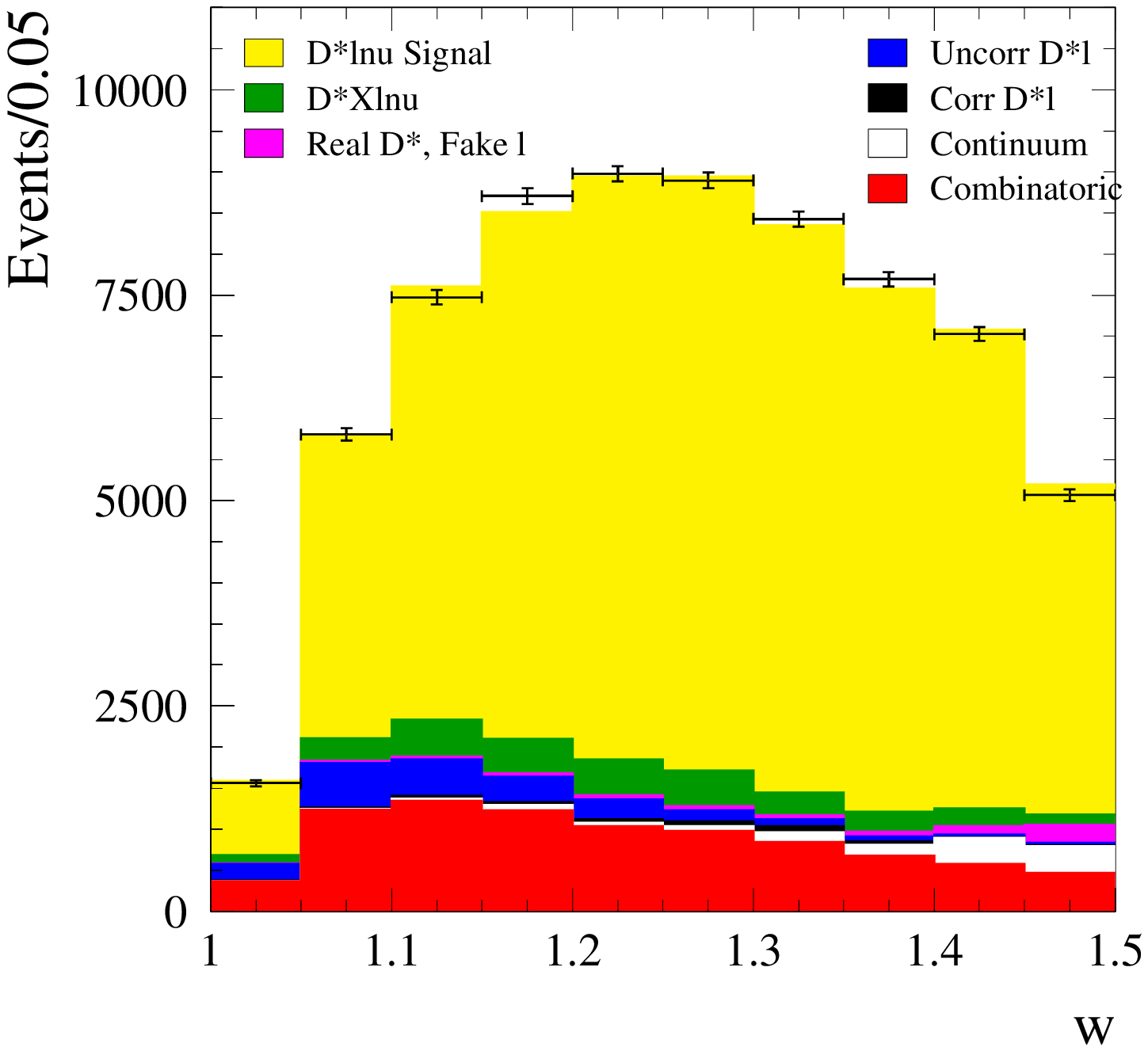}  &
\includegraphics[height=5.5cm,width=5.5cm]{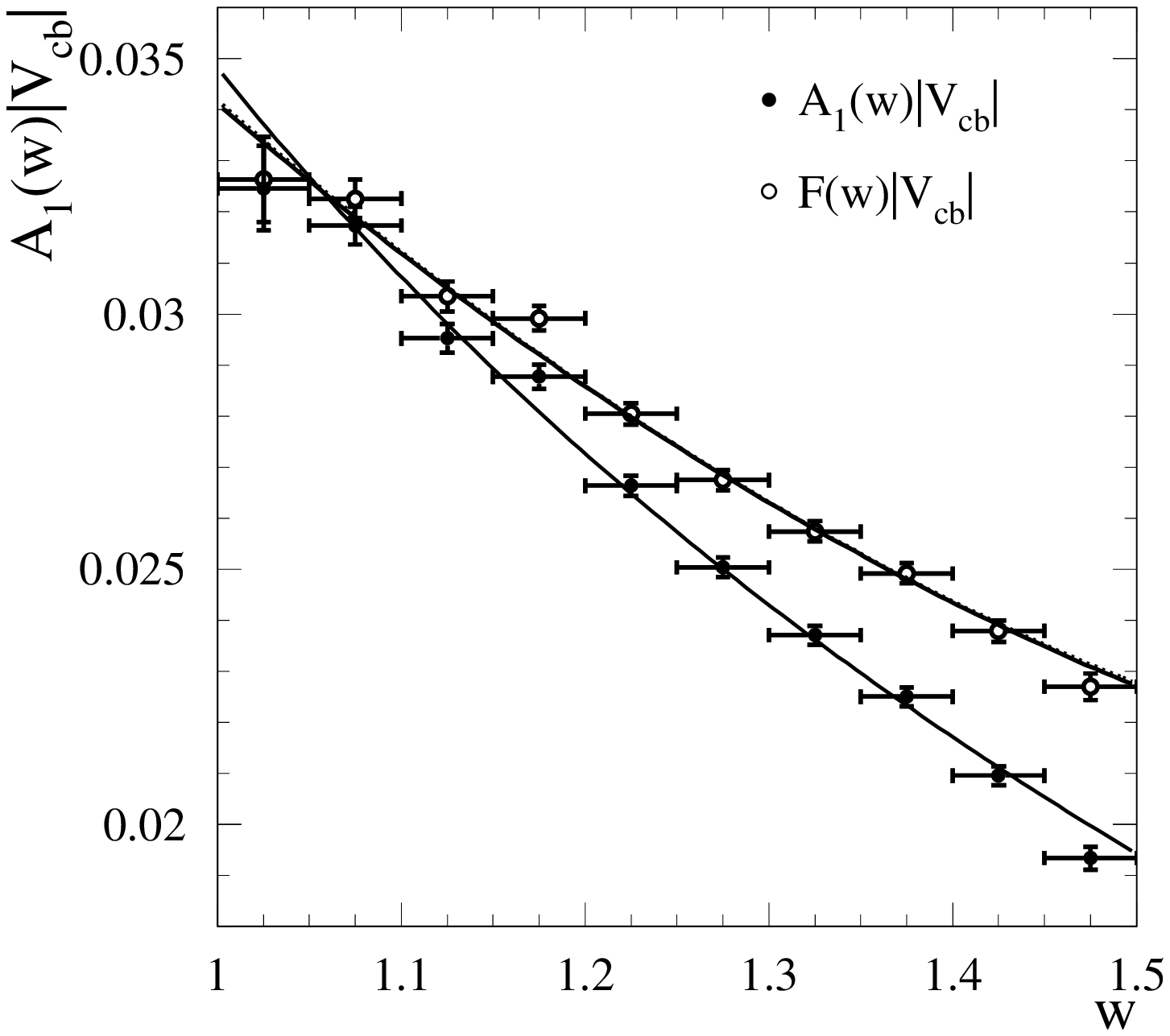}  
\end{tabular}
\vs{-0.7cm}
\caption{Left: Result of the fit (histograms) to the observed $w$ distribution (data points). Right: Form factor dependence on $w$, fits (lines) and data corrected for background and efficiency,  all the terms in the decay rate except for $\Vcb {\cal F}(w)$ for two different parameterizations (see text for details).}
\label{f:fit}
\end{center}
\end{figure}

A measurement of exclusively reconstructed $\Bzb \ra \dsp \ell^- 
\bar{\nu}_{\ell}$ decays rate ($\Gamma$) can be used 
to extract \Vcb~[\cite{Hashimoto}]
by measuring the decay rate $d\Gamma/dw$ for $\om>1$, 
where $w$ is the Lorentz boost of the \dsp\ in the \Bzb\ rest frame,
and extrapolating the rate to the kinematic limit corresponding to \om=1.
The analysis is based on a data sample of  79 (9.6) \invfb\ recorded  on 
(40 \mev below) the \FourS\  resonance.  We select events containing a $D^*$
and an oppositely-charged electron or muon with momentum in the range 
($1.2<p_{\ell}<2.4~\gevc$)\footnote{Momenta are measured in the \FourS\ rest frame, unless explicitely stated otherwise.}.  For each decay candidate we 
compute  the cosine of the angle  between momentum of the \Bzb\ and of the 
$\dsp\ell^-$ pair, $\cos\TBY = \frac{2E_{\Bz} E_{D^*\ell} - M^2_{\Bz} 
- M^2_{D^*\ell} } { 2 p_{\Bz} p_{D^*\ell} }$, that we fit in the range  
$-10 < \cos\TBY < 5$ to determine the signal contribution and the 
normalization of the uncorrelated and \BtoDss\ backgrounds.  
To extract \Vcb, we compare the signal yields to the 
expected differential decay rate $\frac{{\rm d}\Gamma}{{\rm d}\om} = \frac{G^2_F}{48\pi^3} {\cal G}(w) \left[ \Vcb {\cal H}(w) \right]^2$,
where ${\cal G}(w)$ is a known phase space factor and ${\cal H}(w)$ is 
the form factor.  We consider two different parameterization of 
${\cal H}(\om)$. A simple Taylor expansion with three parameters
(the extrapolation ${\cal F}(w=1)\,\Vcb$, the slope $\rho_{\cal{F}}^2$, 
and the curvature $c$) and a parameterization 
with two parameters (the extrapolation $\Vcb \, \Aone(w=1)$ 
and the slope \rha ). The two functions have different slopes, but in 
the limit $\om \ra 1$, we expect $\Aone(1) = {\cal F}(1)$.
We perform a least-squares fit of the sum of the observed signal plus 
background yields to the expected yield in ten bins of \om .  
Figure~\ref{f:fit} compares the observed yield of signal and background events summed over all data samples with the result of the fit and illustrates the extrapolation to $w=1$ for the two form factor parameterizations.  
A major source of uncertainty is 
the reconstruction efficiency of the low-momentum pion from the \dsp\ 
decay. Furthermore, there are several uncertainties related to the 
form factors and their parameterization. \newline
The fit results for the two different parameterizations
of the dependence of the form factors on $w$ give consistent results.  
We adopt the result based on the more recent parameterization 
by Caprini {\it et~ al.}~[\cite{Caprini}] and assign the observed difference 
in the extrapolation to $w \ra 1$ as an additional systematic error. 
Using the recent lattice calculation~[\cite{Hashimoto}], we obtain
$|V_{cb}| = (38.03\pm 0.68 \pm 1.07  {\phantom 1}^{+1.25} _{-1.15} {\phantom 1 
} ^{+1.45} _{-1.25} ) \times 10^{-3}$, where the first error is statistical, 
the second systematic, the third the model uncertainty 
(including the choice of the form factor expansion) and the fourth 
reflects the uncertainty in $\Aone(1)$.

\section{Conclusions and outlook}

\babar~ studies of semileptonic $B$ decays have given a significant 
contribution to the understanding of the theoretical framework used
to describe \btou~ and \btoc~ transitions (OPE and HQET) and 
resulted in a consistent reduction of the error on the \Vub~
and \Vcb~ CKM matrix elements. The \Vub~ inclusive analysis result
is currently the best single measurement with a relative error
on \Vub~ of $\sim$14\% and S/B $\sim$ 1.7 while the exclusive analysis
gave promising preliminary results. The \Vcb~ exclusive analysis
is currently the most precise single measurement and is consistent
with results from Belle and LEP, while the preliminary result
of the moment analysis will soon become public still reducing 
the uncertainty on \Vcb.

\end{document}